\documentclass[%
 reprint,
 amsmath,amssymb,
 aps,prl,
]{revtex4-2}

\usepackage{graphicx}
\usepackage{dcolumn}
\usepackage{bm}

\usepackage{amsmath} 
\usepackage{lmodern}
\usepackage{amssymb}

\begin{document}

\preprint{APS/123-QED}

\title{Tuneable terahertz transitions in zigzag graphene nanoribbons}

\author{R. R. Hartmann}
\affiliation{Physics and Astronomy, University of Exeter, Stocker Road, Exeter EX4 4QL, United Kingdom}
\affiliation{Physics Department, De La Salle University, 2401 Taft Avenue, 0922 Manila, Philippines}
\affiliation{Teracan Ltd, United Kingdom}

\author{M. E. Portnoi}
\email{M.E.Portnoi@exeter.ac.uk}
\affiliation{Physics and Astronomy, University of Exeter, Stocker Road, Exeter EX4 4QL, United Kingdom}
\affiliation{Teracan Ltd, United Kingdom}

\begin{abstract}
We show that a transverse electric field applied to a zigzag graphene nanoribbon opens a band gap whose energy can be tuned through the terahertz (THz) range. The field breaks the reflection symmetry of the ribbon, allowing optical transitions that are forbidden in its absence. The relative strengths of transitions polarized parallel and transverse to the ribbon depend strongly on excitation frequency and become equal at a particular frequency. At this frequency, each helicity selectively excites carriers in only one of the two field-induced edge-state valleys, with the selected valley reversed for the opposite helicity. Since this excitation occurs away from the valley minima, the photoexcited carriers have finite group velocity, giving rise to a circular photogalvanic response. Van Hove singularities enhance radiative recombination at the band edge, providing a route towards compact, tunable THz emitters.
\end{abstract}

\maketitle

The terahertz (THz) region of the electromagnetic spectrum lies between the microwave and infrared frequency ranges and is of considerable interest due to its relevance to a wide range of physical processes and applications, including spectroscopy, imaging, and high-speed communications. However, despite significant efforts, the development of compact, portable and tunable THz sources that operate at room temperature remains a challenge~\cite{dragoman2004terahertz,lee2007searching,leitenstorfer20232023}. Considerable progress has been achieved along the semiconductor route, in particular through quantum cascade lasers (QCLs)~\cite{kohler2002terahertz,khalatpour2021high}. Another route is to exploit low-dimensional carbon-based systems~\cite{hartmann2014terahertz,xie2025carbon}, such as graphene, carbon nanotubes, cyclocarbons~\cite{ng2022tuning}, and polyyne chains~\cite{hartmann2021terahertz,ng2025stark}, as building blocks for THz devices.

Several low-dimensional carbon-based systems face significant challenges for THz applications. For example, the synthesis of carbyne-based systems and cyclocarbons, although advancing, remains at an early stage. Graphene, despite its Dirac-like electronic spectrum and high carrier mobility, lacks an intrinsic band gap, while carbon nanotubes, although extensively studied in the THz context, suffer from suppressed optical emission due to dark excitons, resulting in low quantum efficiency.

Graphene nanoribbons (GNRs), in contrast, provide a particularly attractive platform because quantum confinement opens an energy gap while their planar geometry remains compatible with standard device architectures~\cite{wang2021graphene}. Their quasi-one-dimensional electronic structure gives rise to discrete subbands, with the ribbon edge termination strongly influencing the low-energy spectrum. In particular, zigzag GNRs support localized edge states that form nearly dispersionless bands near the Fermi energy. 

Zigzag GNRs share spectral similarities with armchair carbon nanotubes, and the corresponding band structures become closely matched when the ribbon width is approximately equal to half the nanotube circumference~\cite{white2007hidden}. Many of the nanotube angular momentum modes can be matched to the ribbon transverse modes. However, this matching is not exact for the full spectrum. The topologically protected low-energy crossing of the armchair nanotube bands, formed by extended states, cannot be mapped onto the edge-state bands of the zigzag nanoribbon~\cite{nakada1996edge,brey2006electronic,wakabayashi2010electronic}. 

Optical transitions in carbon nanotubes and GNRs have also been extensively studied, revealing both similarities and differences in their selection rules~\cite{chung2016electronic, saroka2017optical,saroka2018hidden}. In armchair carbon nanotubes, optical transitions are governed by conservation of the quantized circumferential angular momentum, while sublattice symmetry suppresses certain transitions, such as those involving the lowest-energy Dirac bands for light polarized along the nanotube axis. In zigzag GNRs, optical transitions for light polarized along the ribbon axis are governed by parity-related selection rules.

Despite the similarities between armchair nanotubes and zigzag GNRs, striking differences emerge when an electric field is applied in the transverse direction. The effect of a transverse electric field on armchair nanotubes has been extensively studied~\cite{li2003electronic} and leads to subband mixing and a modification of the emission spectrum. However, within the nearest-neighbour $\pi$-orbital tight-binding model, a transverse electric field does not open a gap in an armchair nanotube. Zigzag GNRs, however, exhibit markedly different behavior, namely, the emergence of a band gap that scales linearly with an in-plane electric field applied normal to the ribbon axis~\cite{apel2011energy}. 

In this Letter, we show that an electric field applied transverse to the axis of a zigzag nanoribbon not only opens a gap in the edge-state spectrum, but also strongly alters optical transition strengths and gives rise to transitions that are forbidden in the absence of the field. The associated absorption is also enhanced by the van Hove singularities at the band extrema. The gap varies linearly with the applied field and, for experimentally attainable field strengths, lies within the highly desirable terahertz range.  We further show that tuning the field produces strongly polarization-dependent absorption. We also demonstrate that, for a given applied field, circularly polarized light at a particular excitation frequency can selectively excite carriers near one of the two field-induced edge-state valleys, thereby generating a circular photogalvanic response.

The contrasting response of armchair nanotubes and zigzag GNRs to a transverse electric field reflects a fundamental difference in their transverse boundary conditions and the role of inversion symmetry breaking across the ribbon, which generates an effective mass term. This behavior can be understood in terms of Su–Schrieffer–Heeger (SSH)-type physics used in finite and periodic systems, such as dimerized chains~\cite{su1979solitons,su1980soliton} and rings~\cite{ng2022tuning}, as well as in other topological band structures~\cite{hasan2010colloquium,qi2011topological}. Within this framework, the field opens a gap in the system with open boundaries but not in its periodic counterpart, reflecting the different roles of boundary conditions and edge states in the two geometries.


\begin{figure}
    \centering
\includegraphics[width=\columnwidth]{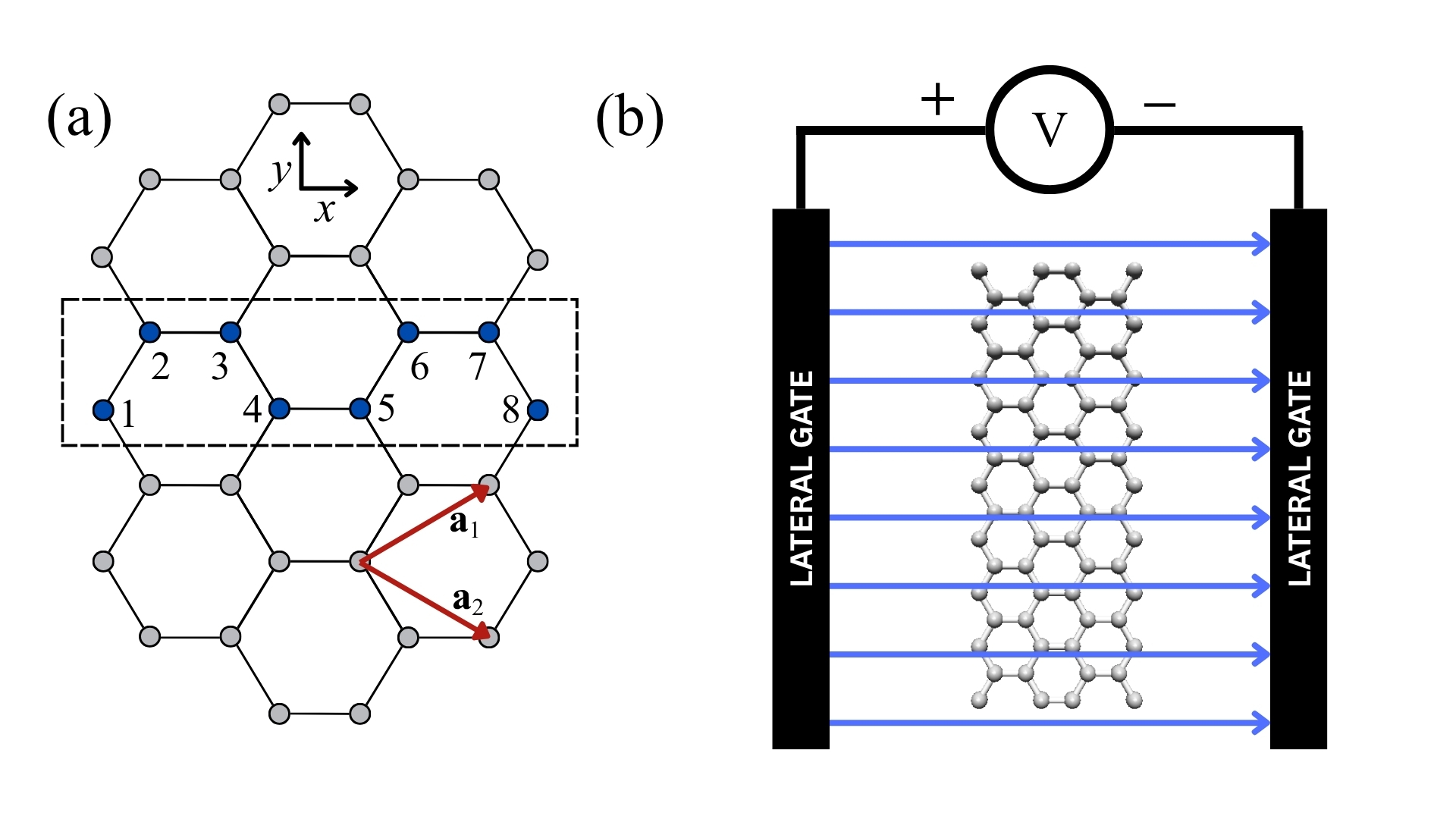}
\caption{(a) The atomic structure of a zigzag graphene nanoribbon, $N$ atoms wide, is shown. Such a nanoribbon is commonly referred to as a $(N/2)$-ZGNR, where $N/2$ is the number of zigzag carbon chains across the ribbon's width. The carbon atoms in the unit cell are numbered, and the primitive lattice vectors of graphene, $\mathbf{a}_1$ and $\mathbf{a}_2$, are indicated, together with the two inequivalent atoms belonging to the A and B sublattices that form the honeycomb lattice. (b) Schematic of the experimental geometry used to apply an in-plane electric field transverse to the nanoribbon axis.}
    \label{fig0}
\end{figure}

In what follows, we employ a nearest-neighbour tight-binding model~\cite{dresselhaus1998physical} to study finite-width zigzag GNRs. This model has proven highly effective in describing the electronic and optical properties of carbon nanostructures, particularly graphene and carbon nanotubes~\cite{hartmann2014terahertz}. The Hamiltonian of a finite-width zigzag GNR, schematically shown in Fig.~\ref{fig0}, with a width of $N$ carbon atoms (where $N$ is an even integer), can be written as
\begin{equation}
\textbf{H}=
\left(\begin{array}{cccccc}
0 & tf\\
tf & 0 & t\\
 & t & 0 & tf\\
 &  & tf & 0 & \ddots\\
 &  &  & \ddots & \ddots & tf\\
 &  &  &  & tf & 0
\end{array}\right)
\label{eq:Ham_zero_field}
\end{equation}
where $t\approx -3$~eV is the hopping integral of graphene, 
$f=2\sin\left(k_{y}a/2\right)$, $k_y$ is the shifted longitudinal wave vector defined by $k_{y}a=\pi-\widetilde{k}_{y}a$, with $\widetilde{k}_{y}$ denoting the original longitudinal wave vector, and $a=a_{cc} \sqrt 3$ with $a_{\rm cc}$ being the carbon–carbon interatomic distance. This Hamiltonian is a finite version of the above-mentioned SSH Hamiltonian, which has recently been applied to finite polyene carbon chains~\cite{hartmann2021terahertz}. In that previous work, $f$ was a positive constant less than $1$ and $t$ played the role of the hopping parameter for triple bonds. 

For a zigzag GNR, depending on the value of $k_y$, the states may be either edge-localized or bulk-like. Edge-state solutions occur for $\left|f\right|<N/\left(N+2\right)$, which approaches $\left|f\right|<1$ in the large-$N$ limit; the remaining solutions are bulk-like. Edge-state solutions are readily obtained in the limiting case $f = 0$, where Eq.~(\ref{eq:Ham_zero_field}) decouples into $(N - 2)/2$ independent dimers and two isolated outer atoms. In this case, the energy spectrum consists of two $\varepsilon = 0$ states associated with the isolated sites, and $\varepsilon = \pm \left|t\right|$ states for each dimer. 

The eigenproblem of the Hamiltonian, Eq.~(\ref{eq:Ham_zero_field}), can be solved using the transfer-matrix method~\cite{molinari1997transfer}, which has been applied to one-dimensional systems~\cite{kerner1954periodic,schmidt1957disordered,hori1957vibration,matsuda1962transfer}, or via continuants, as employed in studies of conjugated $\pi$-electron systems~\cite{lennard1937electronic,coulson1938electronic,mestechkin2005finite,rutherford1948xxv,rutherford1952xvi,muir2003treatise}. Recent advances in the study of tridiagonal matrices~\cite{yueh2005eigenvalues,kouachi2006eigenvalues,da2007characteristic,willms2008analytic,kouachi2008eigenvalues} have also been applied to the study of the optical properties of finite-length polyyne chains~\cite{hartmann2021terahertz,ng2025stark}.

In the presence of an electric field of strength $E$, applied along the $x$-direction (normal to the zigzag GNR axis), the perturbation to the Hamiltonian of Eq.~(\ref{eq:Ham_zero_field}) is $\delta\textbf{H}$, which has only diagonal elements, given by 
\begin{equation*}
\delta H_{i,i}=U_{0}\left[\frac{3}{4}\left(i-\frac{N+1}{2}\right)-\frac{1}{8}\left(-1\right)^{i}\right],
\label{eq:field_peturb}
\end{equation*}
where $U_{0}=eEa_{\rm cc}$. 
In the limits that $N^2\gg1$ and $t^2 \gg \varepsilon^2,U_0^2$, the characteristic equation, $\mathrm{det}\left(\textbf{H}+\delta\textbf{H}-\varepsilon\textbf{I}\right)=0$, can be approximated as a quadratic in $\varepsilon$. Solving this, the edge-state eigenvalues are approximately given by
\begin{equation}
\varepsilon=\pm\sqrt{U_{0}^{2}\Delta_{N}^{2}+t^{2}\kappa_N^2},
\label{eq:spectrum_analytic}
\end{equation}
where
\begin{equation}
\Delta_{N}=1+\frac{3N}{8}+\frac{3}{2\left(f^{2}-1\right)},\qquad\kappa_N=f^{N/2}\left(1-f^2\right).
\nonumber
\end{equation}
Equation~(\ref{eq:spectrum_analytic}) is equivalently obtained as the spectrum of the effective two-band Hamiltonian
\begin{equation} \mathbf{H}_{\mathrm{eff}}=U_{0}\Delta_{N}\sigma_{z}+\left|t\right|\kappa_{N}\sigma_{x},
\label{eq:effective_two_band}
\end{equation}
written in the basis of states localized at the two opposite edges of the ribbon, where $\sigma_x$ and $\sigma_z$ are the Pauli matrices. The derivation and the exact finite-$N$ expressions for $\kappa_N$ and $\Delta_N$ appearing in Eq.~(\ref{eq:effective_two_band}) are given in End Matter, Appendix~A.
\begin{figure}
    \centering
\includegraphics[width=0.8\columnwidth]{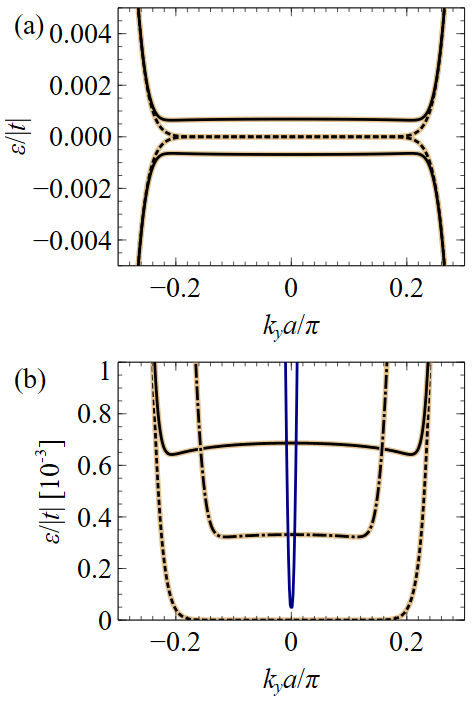}
\caption{(a) Energy spectrum of a 20-ZGNR in the absence (dashed black curve) and presence (solid black curve) of a transverse electric field of strength $10^{6}~\mathrm{V\,m^{-1}}$. (b) Enlarged view of the positive-energy region, showing the spectra of 2-, 10-, and 20-ZGNRs under the same applied field, together with the zero-field spectrum of the 20-ZGNR for reference. For the 10- and 20-ZGNRs, the numerical results (solid and dash-dotted black curves) are compared with the analytical dispersion given by Eq.~(\ref{eq:spectrum_analytic}) (gold curves). The 2-ZGNR spectrum is obtained from the exact analytical result derived in End Matter, Appendix~{B} (dark-blue curve).}
    \label{fig:fig_band}
\end{figure}

In Fig.~\ref{fig:fig_band}, we plot the edge-state bands for an $N=40$ zigzag GNR, obtained numerically and from the analytical expression, Eq.~(\ref{eq:spectrum_analytic}), in the absence of a field (grey line) and for an applied electric field of strength $10^6$~V/m (black line), corresponding to $\left|U_{0}/t\right|=4.73 \times 10^{-5}$. It can be seen from the figure that the applied field opens a gap, and the positive- and negative-energy edge-state bands acquire two additional local minima and two additional local maxima, respectively. For $N>10$, the analytic expression gives excellent agreement with the numerical results, which is lost for smaller matrices.

\begin{figure}
    \centering
\includegraphics[width=0.8\columnwidth]{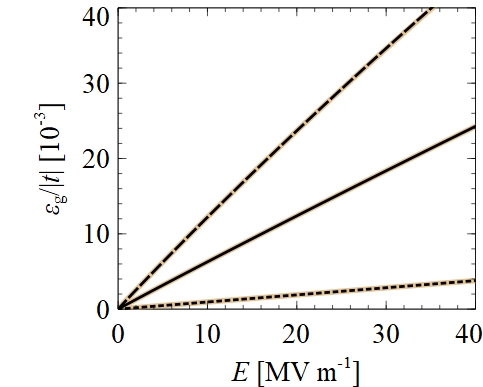}
    \caption{Edge-state gap $\varepsilon_g$ as a function of the transverse electric-field strength. The numerical results are shown in gold. From top to bottom, the black curves correspond to the 20-ZGNR (long-dashed), 10-ZGNR (solid), and 2-ZGNR (short-dashed). The results for the 10- and 20-ZGNRs are obtained from the analytical dispersion in Eq.~(\ref{eq:spectrum_analytic}), while the exact analytical result for the 2-ZGNR is derived in End Matter, Appendix~{B}.
}
\label{fig:gap_vs_N}
\end{figure}

For small electric fields, the edge-state gap, $\varepsilon_g$, scales approximately linearly with $\left|U_{0}\right|$ (see Fig.~\ref{fig:gap_vs_N}), in stark contrast to quasi-metallic zigzag nanotubes, where, below a critical field strength, a transverse electric field (applied perpendicular to the tube axis) opens a gap that scales quadratically with field strength, $\varepsilon_g \approx (e E R)^2/(6\left|t\right|)$, while armchair nanotubes remain metallic since the Dirac crossing is topologically protected~\cite{li2003electronic}. Therefore, zigzag GNRs have a significant advantage over quasi-metallic nanotubes, as the gap can be tuned linearly with the applied field at experimentally accessible strengths, providing practical access to the terahertz regime. The planar nature of GNRs makes them well suited for integration into device architectures, where their electronic properties can be tuned electrostatically. Transverse potentials can be realized either by applying a uniform electric field across the ribbon, for example in a capacitor-like geometry, or via gate electrodes that generate opposite potentials at the boundaries. Such geometries enable direct control of the edge-state spectrum, and, as we show below, provide control over optical transitions between edge states.

Within first-order time-dependent perturbation theory, the transition rate from an initial state $\Psi_i$ with energy $\varepsilon_i$ to a final state $\Psi_f$ with energy $\varepsilon_f$ is given by Fermi's Golden Rule:
\begin{equation}
    \Gamma=\sum_{k_y}\frac{2\pi}{\hbar}\left|\left\langle \psi_{f}\left|\textbf{H}'\right|\psi_{i}\right\rangle \right|^{2}\delta\left(\varepsilon_{f}-\varepsilon_{i}-\hbar\omega\right),
    \label{eq:Fermi_Golden}
\end{equation}
where $\psi_{i}$ and $\psi_{f}$ are the initial and final state wave functions, $\varepsilon_i$ and $\varepsilon_f$ their corresponding energies, and $\textbf{H}'$ is the perturbation due to light with photon energy $\hbar \omega$. Dipole optical transitions in zGNRs have been studied extensively in the absence of a gap-opening field for light polarized along the nanoribbon axis~\cite{saroka2017optical,saroka2018hidden}, where simple tight-binding models show very good agreement with more sophisticated \textit{ab initio} calculations~\cite{payod20202n+}. 

For systems whose dimensions are much smaller than the optical wavelength, such as GNRs subject to light polarized along the transverse ($x$) direction, the optical perturbation may be described using the position operator. 
In this case, $\textbf{H}'_{x} = e\mathcal{E}_{x}x$, where $\boldsymbol{\mathcal{E}}$ is the electric field of light, and $e$ is the elementary charge. However, for light polarized along the ribbon axis ($y$), the optical matrix elements are more naturally evaluated using the velocity operator, $\widehat{v}_{y}=i\left[\boldsymbol{\mathrm{H}},y\right]/\hbar$, or equivalently within the gradient approximation~\cite{blount1962formalisms,johnson1973optical}, so that $\textbf{H}'_{y} = -ie\mathcal{E}_{y} \widehat{v}_{y} / \omega $. For Eq.~(\ref{eq:Ham_zero_field}), the gradient approximation gives $\widehat{v}_{y}=-\left(t/\hbar\right)\partial_{k_{y}}f\left(I_{N/2}\otimes\sigma_{x}\right)$, where $I_{N/2}$ denotes the $(N/2)\times(N/2)$ identity matrix and $\sigma_x$ is the Pauli matrix. 

In the absence of the transverse electric field, the eigenstates have definite parity. Since the velocity operator $\widehat{v}_{y}$ is even under reflection, its matrix element between the opposite-parity valence- and conduction-edge states vanishes. Consequently, the corresponding $y$-polarized transition is forbidden in the absence of the field. In contrast, the position operator $x$ is odd under reflection and therefore couples states of opposite parity, making the corresponding $x$-polarized transition allowed. 

\begin{figure}
    \centering
\includegraphics[width=0.8\columnwidth]{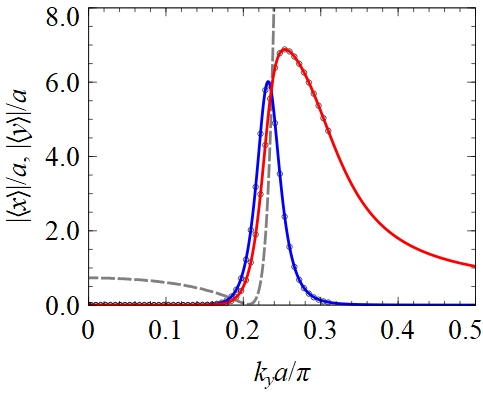}
\caption{Optical matrix elements for transitions across the edge-state gap in a 20-ZGNR as functions of the longitudinal wavevector $k_y$. The red and blue curves show $|\langle x\rangle|/a$ and $|\langle y\rangle|/a$, respectively, where $|\langle x\rangle|=|\langle\psi_f|\hat{x}|\psi_i\rangle|$ and $|\langle y\rangle|=|\langle\psi_f|\hat{v}_y|\psi_i\rangle|/\omega$. These correspond to light polarized transverse and parallel to the ribbon axis, respectively, with $\hbar\omega=\varepsilon_f-\varepsilon_i$. The open circles show the results obtained from the analytical expressions and are plotted over the range in which the states remain edge-localized. The grey dashed line shows the shifted energy spectrum, allowing one to identify the values of $k_y$ corresponding to the strongest transitions.
 }
\label{fig:MOT}
\end{figure}

Within the two-band approximation, the projected position and velocity operators are $\widehat{x}=a_{\mathrm{cc}}\Delta_{N}\sigma_{z}$ and $\widehat{v}_{y}=-\left(\partial_{k_{y}}\mathbf{H}_{\mathrm{eff}}\right)/\hbar$. It should be noted that the negative sign stems from the shift in $k_y$, as described after Eq.~(\ref{eq:Ham_zero_field}). Introducing the mixing angle $\eta$, defined by $\sin\eta=|t\kappa_N|/|\varepsilon|$ and $\cos\eta=U_0\Delta_N/\left|\varepsilon\right|$, the corresponding interband matrix elements are given below (see End Matter, Appendix~{A}):
\begin{equation*}
\left\langle\psi_f\middle|\widehat{x}\middle|\psi_i\right\rangle
=a_{\mathrm{cc}}\Delta_N\sin\eta,
\qquad
\left\langle\psi_f\middle|\widehat{y}\middle|\psi_i\right\rangle
=-\frac{i}{2}\frac{\partial\eta}{\partial k_y}.
\label{eq:two_band_matrix_elements}
\end{equation*}
At zero field, $\eta$ is independent of $k_y$, and hence the $y$-polarized matrix element vanishes. The applied transverse electric field breaks the reflection symmetry of the ribbon. Consequently, the resulting edge states no longer have definite parity, removing the selection rule that forbids the $y$-polarized transition, while the x-polarized transition remains allowed. This contrasts with finite-length carbon chain arrays, where the optical response is dominated by the projection of the electric field along the chain axis~\cite{kucherik2024polarization}. In Fig.~\ref{fig:MOT}, we plot the transverse dipole matrix element
$\left|\left\langle \psi_f \left| \hat{x} \right| \psi_i \right\rangle\right|$ (red) and the effective longitudinal dipole matrix element $\left|\left\langle \psi_{f}\left|\hat{v}_{y}\right|\psi_{i}\right\rangle \right|/\omega$ (blue) as functions of the longitudinal wave vector $k_y$ for a zigzag GNR with a width of $N=40$ atoms (20-ZGNR). As shown in Fig.~\ref{fig:MOT}, both matrix elements are strongly suppressed near the centre of the Brillouin zone and exhibit pronounced peaks close to the band edge. Away from these peaks, the transverse matrix element approaches a finite value, whereas the matrix element for light polarized along the ribbon axis rapidly vanishes. At different excitation frequencies, the optical response may be dominated by different polarization channels.

The polarization anisotropy is quantified by the degree of alignment of the photo-induced dipole with the incident light polarization, which is determined by the relative strengths of the longitudinal and transverse transition matrix elements. 
For normal incidence (the results can be readily generalized to arbitrary angles of incidence by projecting the light polarization onto the ribbon plane), the polarization vector $\boldsymbol{e}$ can be written as
$
\boldsymbol{e}=\left(\cos\theta,e^{i\phi}\sin\theta\right)$. Here, circularly polarized light corresponds to $\theta=\pi/4$, with right- and left-handed circular polarization defined by $\phi=-\pi/2$ and $+\pi/2$, respectively, whereas for linearly polarized light $\phi=0$ and $\theta$ is the angle between the polarization vector and the $x$ axis. For an arbitrary polarization of light incident normal to the zGNR plane, the probability of dipole transitions is proportional to 
\begin{equation}
\left|\left\langle \psi_{f}\right|\boldsymbol{r}\cdot\boldsymbol{e}\left|\psi_{i}\right\rangle \right|^{2}=\frac{S_{0}}{2}\left[1+\alpha\cos\left(2\theta+\delta\right)\right]
\label{eq:dipole_eqn}
\end{equation}
where
\begin{equation*}
\begin{aligned}
S_0 &=
\left\langle x\right\rangle ^{2}+\left\langle y\right\rangle ^{2},
\quad
\alpha=\sqrt{1-\frac{4\left\langle x\right\rangle ^{2}\left\langle y\right\rangle ^{2}\cos^{2}\left(\phi\right)}{S_{0}^{2}}}
\\
\alpha\cos\left(\delta\right) &=
\frac{\left\langle x\right\rangle^{2}
-\left\langle y\right\rangle^{2}}
{S_0},
\quad
\alpha\sin\left(\delta\right) =
\frac{2\left\langle x\right\rangle
\left\langle y\right\rangle
\sin\left(\phi\right)}
{S_0}.
\end{aligned}
\end{equation*}
Here, in the phase convention used, $\langle x\rangle=\langle\psi_f|\widehat{x}|\psi_i\rangle$ and
$\langle y\rangle=-\langle\psi_f|\widehat{v}_y|\psi_i\rangle/\omega$
are real. The physical longitudinal dipole matrix element is purely
imaginary, $\langle\psi_f|\widehat{y}|\psi_i\rangle=i\langle y\rangle$;
hence, $\langle y\rangle$ represents its imaginary part.

\begin{figure}
    \centering
    \includegraphics[width=0.8\linewidth]{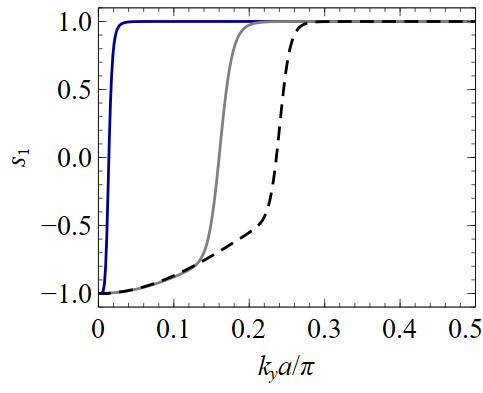}
    \caption{Normalized Stokes parameter $s_1$ for linearly polarized excitation as a function of the longitudinal wavevector $k_y$. The leftmost blue, middle grey, and rightmost dashed black curves correspond to the 2-, 10-, and 20-ZGNRs, respectively. The limits $s_1=1$ and $s_1=-1$ correspond to purely $x$- and $y$-polarized transitions, respectively. The transverse electric-field strength is $1~\mathrm{MV\,m^{-1}}$.}
    \label{fig:alpha}
\end{figure}

\begin{figure}
    \centering
\includegraphics[width=\columnwidth]{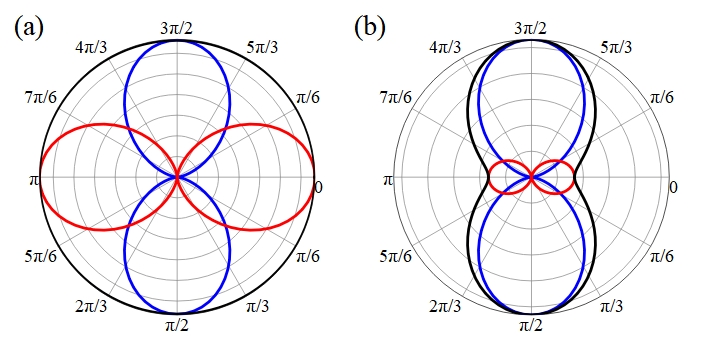}
    \caption{Polar plots of the squared optical matrix-element contributions
$\left|\left\langle\psi_f\middle|\hat{x}\middle|\psi_i\right\rangle\right|^2\cos^2\theta$
(red) and
$\left|\left\langle\psi_f\middle|\hat{v}_y\middle|\psi_i\right\rangle/\omega\right|^2\sin^2\theta$
(blue), corresponding to light polarized transverse and parallel to the ribbon axis, respectively, together with their sum (black). The polarization angle $\theta$ is measured from the $x$-axis, and the linearly polarized light is incident normal to the ribbon plane. Results are shown for a 20-ZGNR under a transverse electric field of $1~\mathrm{MV\,m^{-1}}$: (a) at the frequency for which the transverse and longitudinal matrix elements have equal magnitudes and (b) at the band-edge transition.}
\label{fig:polar_plot}
\end{figure}

For linearly polarized excitation, the states with opposite signs of $k_y$ are equally populated, and the considered interband dipole transitions are characterized by the normalized Stokes parameter $s_1$, given by
\begin{equation*}
s_1=\frac{\left\langle x\right\rangle ^{2}-\left\langle y\right\rangle ^{2}}
{\left\langle x\right\rangle ^{2}+\left\langle y\right\rangle ^{2}}.
\end{equation*}
For linearly polarized light, the parameter $\alpha$ introduced earlier satisfies
$\alpha = |s_1|.$ In Fig.~\ref{fig:alpha}, we plot $s_1$ for three different nanoribbon widths. The positive sign of $s_1$ observed at high frequencies, together with its tendency to approach unity, indicates that transitions induced by light polarized along the $x$ direction, perpendicular to the ribbon, dominate over those induced by light polarized along the $y$ direction, parallel to the ribbon. At lower excitation energies, $s_1$ passes through zero, corresponding to a complete suppression of the polarization anisotropy (see Fig.~\ref{fig:polar_plot}(a)). As can be seen from Fig.~\ref{fig:MOT}, this occurs when the longitudinal and transverse transition matrix elements become equal in magnitude. Consequently, the parameter $\alpha$ vanishes at the corresponding excitation energy. At the band edge and for the applied field considered here, transitions induced by light polarized along the $y$ direction dominate, producing the strongly anisotropic optical response shown in Fig.~\ref{fig:polar_plot}(b).

More generally, varying the applied field modifies the band structure and changes the relative magnitudes of $\left\langle x\right\rangle$ and $\left\langle y\right\rangle$, thereby tuning the polarization anisotropy. In contrast to linear carbon chains \cite{kucherik2024polarization}, whose optical response is intrinsically polarized along the chain axis, zigzag GNRs exhibit electric-field-tunable optical anisotropy.

\begin{figure}
    \centering
    \includegraphics[width=0.8\linewidth]{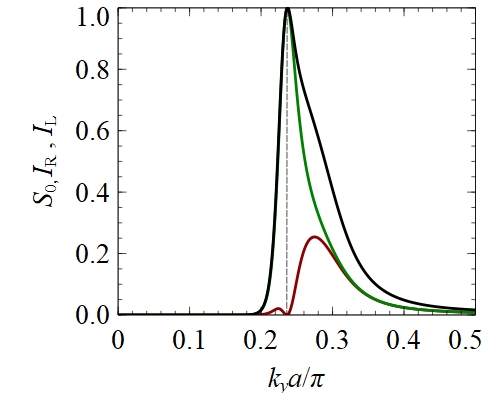}
    \caption{Total transition strength $S_0$ (black) and the right- and left-handed helicity-resolved transition strengths $I_{\mathrm{R}}$ (red) and $I_{\mathrm{L}}$ (green), respectively, as functions of the longitudinal wavevector $k_y$ for a 20-ZGNR. The vertical dashed line marks the point at which $|\langle x\rangle|=|\langle y\rangle|$, where $I_{\mathrm{R}}$ vanishes and perfect helicity selectivity is obtained. All transition strengths are normalized to the maximum value of $S_0$. The transverse electric-field strength is $1~\mathrm{MV\,m^{-1}}$.}
    \label{fig:Stokes}
\end{figure}

In the presence of an applied field, the edge-state dispersion exhibits two minima at opposite longitudinal momenta, which we refer to as field-induced edge-state valleys. These valleys couple differently to the two helicities of circularly polarized light. For right- and left-handed circularly polarized light, Eq.~(\ref{eq:dipole_eqn}) becomes
\begin{equation*}
I_{R,L}
=\frac{\left(\left\langle x\right\rangle \pm \left\langle y\right\rangle \right)^{2}}{2},
\end{equation*}
where the upper and lower signs correspond to right- and left-handed circular polarization, respectively. For states with the opposite sign of $k_y$, i.e. in the region associated with the opposite minimum, the transverse transition matrix element changes sign relative to the longitudinal matrix element, interchanging the right- and left-handed selection rules. Consequently, at the excitation frequency for which $\left|\left\langle x\right\rangle\right|=\left|\left\langle y\right\rangle\right|$, each helicity selectively excites carriers in the region associated with only one of the two minima, with the selected region reversed upon switching the helicity. This helicity selectivity is illustrated in Fig.~\ref{fig:Stokes}, where one of the helicity-resolved transition strengths vanishes while the other remains finite. For a 2-ZGNR, perfect helicity selectivity occurs at $\hbar\omega\approx2\left|t\right|\left(\sqrt{3}\left|U_{0}/t\right|\right)^{2/3}$ in the weak-field limit (see End Matter, Appendix~{B}). Perfectly selective excitation by circularly polarized light occurs away from the band minima, where the carriers have finite group velocity. This selective excitation can therefore generate a net current, giving rise to a circular photogalvanic response. This is in stark contrast to gapped graphene, where complete valley polarization under circularly polarized excitation occurs at the band extrema and the photoexcited carriers consequently have zero group velocity~\cite{hartmann2019interband}.

Recent developments in snapshot polarization imaging allow the full Stokes vector to be determined without sequential polarization analysis~\cite{wang2026stokes}. The electrically tunable anisotropy and helicity selectivity predicted here suggest that arrays of independently gated zGNR elements, configured to analyse complementary linear and circular polarization channels, could enable full-Stokes detection in the THz range.

\begin{figure}
    \centering
\includegraphics[width=\columnwidth]{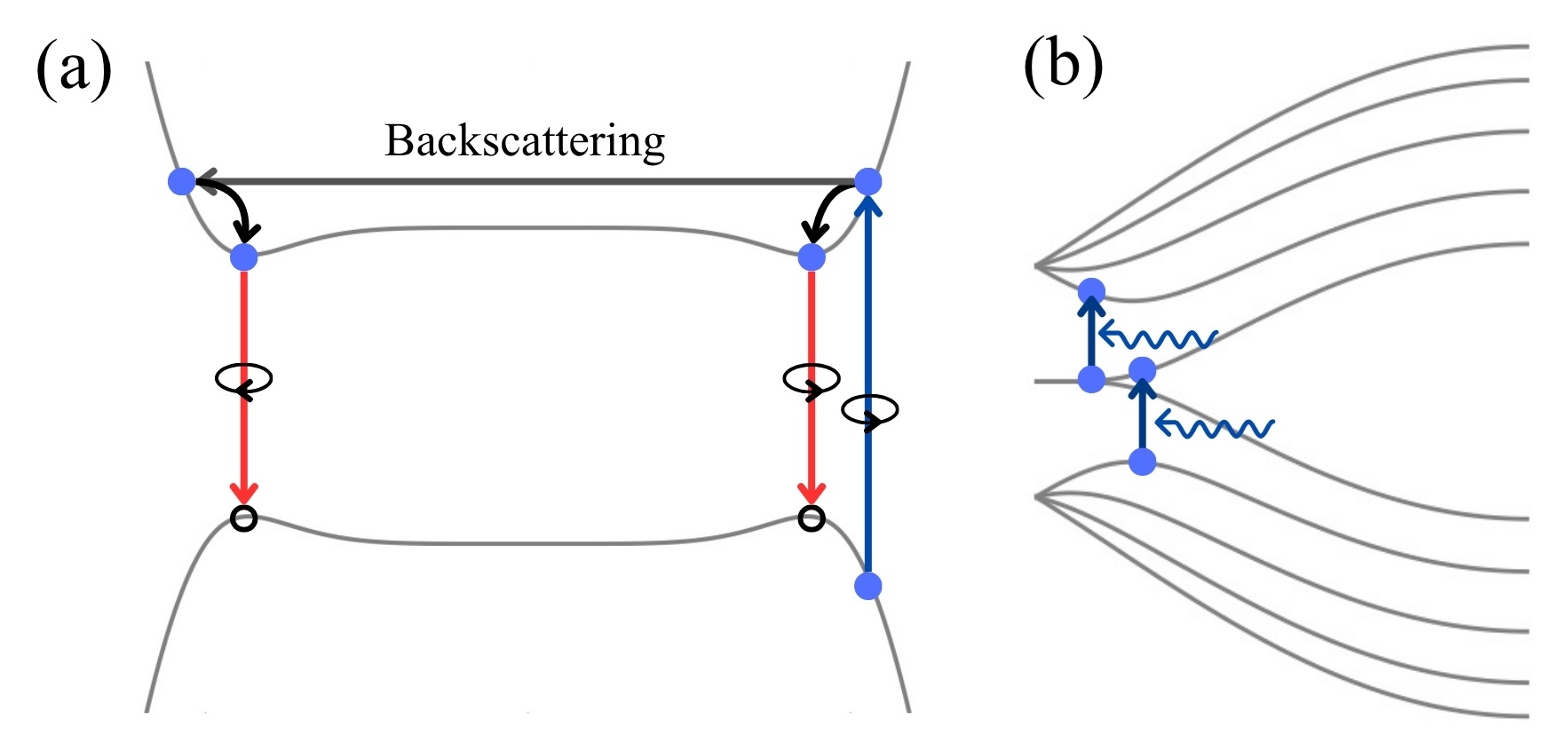}
    \caption{Panel (a) shows a schematic of the proposed experiment, in which circularly polarized THz excitation is used to study the role of backscattering. Panel (b) shows a possible optical excitation scheme that produces a population inversion between states separated by THz frequencies.
    }
    \label{fig:scheme}
\end{figure}

Helicity-resolved band-edge photoluminescence provides a useful tool for studying relaxation in zGNRs. By pumping with circularly polarized light of a given helicity, carriers can be selectively excited near one of the two minima, as illustrated in Fig.~\ref{fig:scheme}(a). After relaxation to the band edge, emission with the corresponding helicity indicates that the carriers remained in the initially populated minimum, whereas emission dominated by the opposite helicity indicates scattering to the other minimum. The degree of circular polarization of the emitted light therefore provides a measure of how strongly the initial momentum population is retained.

\begin{figure}
    \centering
\includegraphics[width=0.8\columnwidth]{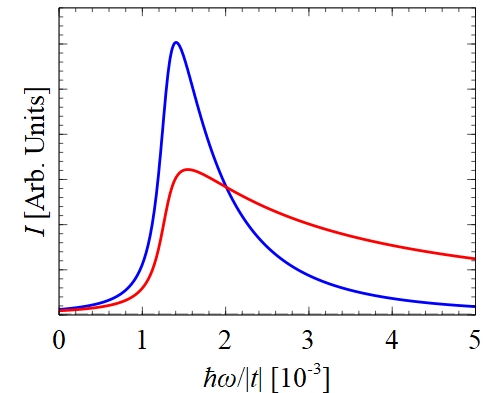}
    \caption{Polarization-resolved emission spectra for transitions between the edge states of a 20-ZGNR ($N=40$ atoms) subjected to a transverse electric field of strength $E=1~\mathrm{MV\,m^{-1}}$. The red and blue curves correspond to emission polarized along the $x$ direction, transverse to the ribbon axis, and the $y$ direction, parallel to the ribbon axis, respectively. }
    \label{fig:emission}
\end{figure}

Beyond its polarization dependence, the emission spectrum is also governed by the density of states. The van Hove singularities in the edge-state bands shown in Fig.~\ref{fig:fig_band} therefore produce pronounced peaks in the broadened emission spectrum shown in Fig.~\ref{fig:emission}. The spectrum was calculated by replacing the delta function in
Eq.~(\ref{eq:Fermi_Golden}) with a Lorentzian of half-width at half-maximum of
$1.5\times10^{-4}|t|=0.45~\mathrm{meV}$.


We also consider the effect of an ac transverse electric field (see End Matter, Appendix~{C}). In the low-energy region near $k_y=0$, corresponding to THz transitions, an ac field does not open a gap at the original band crossing and therefore does not produce a conventional metal-dielectric transition. Nevertheless, the periodic drive modifies the optical transitions between the dressed edge states. The $x$-polarized matrix element remains finite near $k_y=0$, whereas the $y$-polarized matrix element vanishes at $k_y=0$ but becomes finite away from this point. At larger wavevectors, near the photon resonance and outside the THz regime considered here, the periodic drive produces Floquet sidebands and avoided crossings. The resulting dynamical gaps are controlled by the field amplitude and are accompanied by a redistribution of the transition strength between the dressed states, producing strongly polarization-dependent optical transitions.

THz emission may be achieved via optical pumping with polarized light. A wide variety of optical transitions are possible for both polarizations; for simplicity, only two are shown in Fig.~\ref{fig:scheme}(b): one that creates a hole in the lower-energy edge-state band, and another that promotes an electron into the upper edge-state band. Following excitation, both electrons and holes undergo nonradiative relaxation toward the edge-state band extrema, where radiative recombination gives rise to THz emission. The emission is strongly enhanced near the band extrema due to the presence of van Hove singularities in the joint density of states, while its frequency is directly controlled by the applied electric field through the field-dependent gap. Arrays of zigzag GNRs, placed within a suitably designed cavity, could serve as the active element of a coherent THz emitter, forming the basis of a new class of electrically tunable THz sources.

In this work, we neglect many-body effects, which may modify quantitative details but are not expected to alter our main results. In particular, there is the fascinating effect of spin-polarized edge states~\cite{jung2009interedge,magda2014room}, which may give rise to spin currents. In addition, quasi-one-dimensional systems are known to exhibit strongly enhanced excitonic (electron–hole attraction) effects. These effects are expected to be particularly important for the “heavy” quasiparticles associated with the flat portions of the energy dispersion. However, the strongest optical transitions correspond to the steeply rising portions of the energy spectrum, where excitonic effects are expected to be less significant because the photoexcited quasiparticles have large and opposite momenta.

Zigzag GNRs provide a route towards tunable THz optoelectronics. A transverse electric field opens a gap in the otherwise gapless edge-state spectrum, with the gap scaling linearly with the field strength. By breaking the reflection symmetry of the ribbon, the field also enables otherwise forbidden $y$-polarized transitions between edge states. The relative strengths of the longitudinal and transverse transition matrix elements depend strongly on both the excitation frequency and the applied field, giving rise to electrically tunable polarization anisotropy. At particular excitation energies, circularly polarized light can selectively populate carriers near one of the two band minima, while helicity-resolved emission provides a means of probing scattering between them. Van Hove singularities in the edge-state density of states further enhance the emission. Optical pumping followed by carrier relaxation therefore provides a viable mechanism for generating THz radiation whose frequency and polarization can be controlled electrically. These results establish zGNRs as a promising platform for compact, tunable THz sources and as a model system for exploring edge-state physics in low-dimensional Dirac materials.

\begin{acknowledgments}
The authors thank Eros Mariani for helpful discussions. This work was supported by the EU Horizon Europe MSCA Staff Exchange project HERMES (grant agreement No 101236439) and by the UK EPSRC grant EP/Y021339/1.
\end{acknowledgments}




\bibliography{refs}

\section*{End Matter}

\appendix
\setcounter{equation}{0}
\renewcommand{\theequation}{A\arabic{equation}}

\section*{Appendix A: Two-band approximation}\label{app:two_band}
For the matrix Hamiltonian defined in Eq.~(\ref{eq:Ham_zero_field}), in the absence of an applied electric field, the characteristic polynomial in dimensionless units is given by
\begin{equation}
\det\left(\widetilde{\mathbf{H}}-\widetilde{\varepsilon}\mathbf{I}\right)=f^{M-1}\left(f\mathcal{U}_{M}+\mathcal{U}_{M-1}\right),
\nonumber
\end{equation}
where $\widetilde{\mathbf{H}}=\mathbf{H}/\left|t\right|$, $\widetilde{\varepsilon}=\varepsilon/\left|t\right|$, $\mathcal{U}_{M}=\mathcal{U}_{M}\left(\cos\rho\right)$ denotes the Chebyshev polynomial of the second kind, $M=N/2$, and $\cos\rho=\left(\widetilde{\varepsilon}^{2}-1-f^{2}\right)/2f$. For edge states, $\rho=\pi+i\varphi$ for $f>0$ and $\rho=i\varphi$ for $f<0$, where $\varphi$  is real. Therefore, the edge-state condition is
\begin{equation}
\left|f\right|=\frac{\sinh\left(M\varphi\right)}{\sinh\left[\left(M+1\right)\varphi\right]}, 
\nonumber
\end{equation} 
with the edge-state threshold occurring at $\left|f\right|=M/\left(M+1\right)$. It should be noted that, for extended states, the condition becomes $f=-\sin\left(M\rho\right)/\sin\left[\left(M+1\right)\rho\right]$, $\rho\in\mathbb{R}$. Using the edge-state condition to eliminate $f$, the edge-state energies are
\begin{equation}
\widetilde{\varepsilon}=\pm\left|\kappa_{N}\right|,\qquad\kappa_{N}=\frac{s^{M}\sinh\left(\varphi\right)}{\sinh\left[\left(M+1\right)\varphi\right]},
\label{eq:Kappa_N_full}
\end{equation}
where $s=\mathrm{sgn\left(f\right)}$, and the corresponding edge-state wavefunctions $\psi=\left(\psi_{1},\psi_{2},\ldots \psi_{N}\right)^{\mathrm{T}}$ have components
\begin{equation}
\psi_{j}=
\begin{cases}
\pm s(-s)^{l}\mathcal{N}
\sinh\left[(M-l+1)\varphi\right],
& j=2l-1,\\[6pt]
(-s)^{l}\mathcal{N}
\sinh\left(l\varphi\right),
& j=2l.
\end{cases}
\nonumber
\end{equation}
Here, the plus and minus signs correspond to the valence and conduction bands, respectively, $\mathcal{N}$ is the normalization constant, and $l=1,2,\ldots,M$. 

For the dimensionless field perturbation $\widetilde{U}=\widetilde{U}_{0}x/a_{\mathrm{CC}}$ (where $\widetilde{U}_{0}=U_{0}/|t|$), the matrix element is $\left\langle\psi_{f}\right|\widetilde{U}\left|\psi_{i}\right\rangle\equiv\widetilde{U}_{0}\Delta_{N}$, where
\begin{equation}
    \Delta_{N}=
1+\frac{3N}{8}+\frac{3}{8}\left[\frac{4-\left(N+2\right)^{2}\kappa_{N}^{2}}{f^{2}-1+\left(N+1\right)\kappa_{N}^{2}}\right]
\label{eq:Delta_N_full}
\end{equation}
Treating the electric field perturbatively gives the approximate edge-state dispersion
\begin{equation}
\widetilde{\varepsilon}\approx\pm\sqrt{\widetilde{U}_{0}^{2}\Delta_{N}^{2}+\kappa_{N}^{2}}.
\label{eq:Energy_full}
\nonumber
\end{equation}
In the wide-ribbon limit, Eqs.~(\ref{eq:Kappa_N_full}) and~(\ref{eq:Delta_N_full}) become
\begin{equation}
\kappa_{N}\approx f^{N/2}\left(1-f^{2}\right),
\quad
\Delta_{N}\approx
1+\frac{3N}{8}+\frac{3}{2\left(f^{2}-1\right)}.
\nonumber
\end{equation}

Near the band edge, the resulting dispersion can be represented by the two-band effective Hamiltonian given by Eq.~(\ref{eq:effective_two_band}) of the main text:
\begin{equation}
\widetilde{\mathbf{H}}_{\mathrm{eff}}=\widetilde{U}_{0}\Delta_{N}\sigma_{z}+\kappa_{N}\sigma_{x}.
\nonumber
\end{equation}
Within this two-band description, the position operator is $\hat{\boldsymbol{x}}=a_{\mathrm{cc}}\Delta_{N}\sigma_{z}$, while the velocity operator along the ribbon is $\hat{v}_{y}=\left(t/\hbar\right)\partial_{k_{y}}\widetilde{\mathbf{H}}_{\mathrm{eff}}$. The corresponding interband position matrix elements can be expressed in terms of the mixing angle $\eta$ as
\begin{equation}
\left\langle \psi_{f}\right|\widehat{x}\left|\psi_{i}\right\rangle =a_{\mathrm{cc}}\Delta_{N}\sin\left(\eta\right),
\;
\left\langle \psi_{f}\right|\widehat{y}\left|\psi_{i}\right\rangle \equiv i\left\langle y\right\rangle =-\frac{i}{2}\frac{\partial\eta}{\partial k_{y}},
\nonumber
\end{equation}
where $\sin\eta=\left|\kappa_{N}/\widetilde{\varepsilon}\right|$ and $\cos\eta=\widetilde{U}_{0}\Delta_{N}/\left|\widetilde{\varepsilon}\right|$. Hence, the required $\widetilde{U}_{0}$ for $\left|\left\langle x\right\rangle \right|=\left|\left\langle y\right\rangle \right|$ is given by
\begin{equation}
\left|\widetilde{U}_{0}\right|=\frac{2\left|\kappa_{N}\right|}{\sqrt{\left[\partial_{k_{y}'}\ln\!\left|\frac{\kappa_{N}}{\Delta_{N}}\right|\right]^{2}-4\Delta_{N}^{2}}},
\nonumber
\end{equation}
where $k_y'=a_{\rm CC}k_y$.

\setcounter{equation}{0}
\renewcommand{\theequation}{B\arabic{equation}}
\section*{Appendix B: Electronic and optical properties of the 2-ZGNR}
\label{app:2ZGNR}
For a 2-ZGNR, the eigenvalues and eigenvectors can be obtained fully analytically. We continue to use dimensionless energies, denoted by a tilde and measured in units of $|t|$. The tight-binding Hamiltonian of a 2-ZGNR subject to a transverse electric field is given by
\begin{equation}
\boldsymbol{\mathrm{H}}=\left(\begin{array}{cccc}
-U_{0} & tf & 0 & 0\\
tf & -\frac{1}{2}U_{0} & t & 0\\
0 & t & \frac{1}{2}U_{0} & tf\\
0 & 0 & tf & U_{0}
\end{array}\right).
\label{eq:Ham_4by4}
\end{equation}
Dividing Eq.~(\ref{eq:Ham_4by4}) by $|t|$, we introduce the dimensionless quantities
$\widetilde{\boldsymbol{\mathrm{H}}}=\boldsymbol{\mathrm{H}}/|t|$, $\widetilde{U}_0=U_0/|t|$, and
$\widetilde{\varepsilon}=\varepsilon/|t|$. The corresponding dimensionless eigenvalues are
\begin{equation}
\widetilde{\varepsilon}=\pm\sqrt{\frac{1+\frac{5}{4}\widetilde{U}_{0}^{2}+2f^{2}+\gamma}{2}},
\label{eq:4by4_eigen}
\end{equation}
where 
\begin{equation}
\gamma=s_{\gamma}\sqrt{\left(1-\frac{3}{4}\widetilde{U}_{0}^{2}\right)^{2}+4f^{2}\left(1+\frac{9}{4}\widetilde{U}_{0}^{2}\right)},
\nonumber
\end{equation}
where $s_{\gamma}=\pm1$. When $s_{\gamma}=-1$, Eq.~(\ref{eq:4by4_eigen}) yields a pair of bands separated by the electric-field-induced gap,
\begin{equation}
\widetilde{\varepsilon}_{g}=\frac{2|\widetilde{U}_{0}|}
{\sqrt{1+\frac{9}{4}\widetilde{U}_{0}^{2}}}.
\nonumber
\end{equation}
Taking $\widetilde{\varepsilon}>0$, the normalized eigenstates of the
low-energy conduction and valence bands are
\begin{equation}
\psi_{c}=\frac{1}{\sqrt{\mathcal{N}_{4}}}\begin{pmatrix}-\dfrac{Af}{\widetilde{\varepsilon}+\widetilde{U}_{0}}\\
A\\
\widetilde{U}_{0}-\widetilde{\varepsilon}\\
f
\end{pmatrix},
\qquad
\psi_{v}=\frac{1}{\sqrt{\mathcal{N}_{4}}}\begin{pmatrix}f\\
\widetilde{\varepsilon}-\widetilde{U}_{0}\\
A\\
\dfrac{Af}{\widetilde{\varepsilon}+\widetilde{U}_{0}}
\end{pmatrix}
\label{eq:low_energy_eigenstates}
\nonumber
\end{equation}
where $A=(\widetilde{\varepsilon}-\widetilde{U}_{0})
(\widetilde{\varepsilon}-\widetilde{U}_{0}/2)-f^{2},$ and the normalization constant is
\begin{equation}
\mathcal{N}_{4}=\left(\widetilde{\varepsilon}-\widetilde{U}_{0}\right)^{2}+f^{2}+\left[1+\frac{f^{2}}{\left(\widetilde{\varepsilon}+\widetilde{U}_{0}\right)^{2}}\right]A^{2}.
\nonumber
\end{equation}
Using these eigenstates, the transverse position matrix element
$\langle x\rangle=\langle\psi_c|\hat{x}|\psi_v\rangle$ and the imaginary
part of the longitudinal position matrix element
$\langle y\rangle=\operatorname{Im}\langle\psi_c|\hat{y}|\psi_v\rangle$
are
\begin{equation}
\left\langle x\right\rangle =\frac{aA\left(2f^{2}-\widetilde{\varepsilon}^{2}+\widetilde{U}_{0}^{2}\right)}{\sqrt{3}\,\mathcal{N}_{4}\left(\widetilde{\varepsilon}+\widetilde{U}_{0}\right)},
\nonumber
\end{equation}
\begin{equation}  
\left\langle y\right\rangle  =-\frac{2aA\widetilde{U}_{0}\sin\left(k_{y}a\right)}{\mathcal{N}_{4}\widetilde{\varepsilon}\left(\widetilde{\varepsilon}+\widetilde{U}_{0}\right)}.
\nonumber
\end{equation}
In the regime $\left|\widetilde{U}_{0}\right| \ll \left|k_{y}\right| a$ and $\left|\widetilde{U}_{0}\right| \ll 1$, the transition matrix elements to first order in $\widetilde{U}_0$ are
\begin{equation}
\left\langle x\right\rangle =-\frac{a}{4\sqrt{3}}\left(1+\frac{3}{\sqrt{1+4f^{2}}}\right),
\nonumber
\end{equation}
\begin{equation}
\left\langle y\right\rangle =\frac{a\widetilde{U}_{0}\sin\left(k_{y}a\right)}{\widetilde{\varepsilon}^{2}\sqrt{1+4f^{2}}},
\nonumber
\end{equation}
where $\widetilde{\varepsilon}\approx(\sqrt{1+4f^{2}}-1)/2$.
The two transition matrix elements are equal in magnitude at the characteristic photon energy and wave vector
\begin{equation}
\hbar\omega\approx2|t|\left(\sqrt{3}\,|\widetilde{U}_{0}|\right)^{2/3},\qquad
|k_{y}|\approx\frac{1}{a}\left(\sqrt{3}\,|\widetilde{U}_{0}|\right)^{1/3}.
\nonumber
\end{equation}

\section*{Appendix C: Floquet response to a transverse ac electric field}
\label{app:ac_field}
Here, we consider the effect of a transverse ac electric field, $E\left(t\right)=E_{0}\cos\left(\Omega t\right)$. Let $\left|c\right\rangle$  and $\left|v\right\rangle$  denote the zero-field conduction and valence edge states of the ribbon, with energies $\varepsilon$  and $-\varepsilon$ , respectively. Within the two-band approximation, the driven Hamiltonian is
\begin{equation}
\boldsymbol{\mathrm{H}}\left(t\right)=\varepsilon\sigma_{z}+eE_{\mathrm{0}}x_{cv}\cos\left(\Omega t\right)\sigma_{x},
\nonumber
\end{equation}
where $x_{cv}=\left\langle c\right|\hat{x}\left|v\right\rangle$ , chosen to be real. Writing the periodic part of a Floquet state as
\begin{equation}
 \left|u\left(t\right)\right\rangle =\sum_{n}e^{in\Omega t}\left|u_{n}\right\rangle,
\nonumber
\end{equation}
 we denote the $n$th Floquet replica of the zero-field state $\left|\beta\right\rangle$, with $\beta=c,v,$ by $\left|\beta,n\right\rangle$ ~\cite{shirley1965solution}. Upon truncating the Floquet expansion to the three replicas $n=-1,0,1$ and choosing the ordered basis $\left(c_{-1},v_{0},c_{1},v_{-1},c_{0},v_{1}\right)$, the Floquet Hamiltonian separates into two independent blocks, $\left(c_{-1},v_{0},c_{1}\right)$ and $\left(v_{-1},c_{0},v_{1}\right)$. The two $3\times3$ Floquet blocks are
\begin{equation}
\boldsymbol{\mathrm{H}}^{u,l}=\left(\begin{array}{ccc}
-\hbar\Omega & X_{cv} & 0\\
X_{cv} & 0 & X_{cv}\\
0 & X_{cv} & \hbar\Omega
\end{array}\right)\pm\varepsilon\left(\begin{array}{ccc}
1 & 0 & 0\\
0 & -1 & 0\\
0 & 0 & 1
\end{array}\right),
\nonumber
\end{equation}
where $X_{cv}=eE_{0}x_{cv}/2$ and the upper and lower signs correspond to the upper and lower blocks, respectively. At $\varepsilon=0$, both blocks possess a zero-quasienergy state. First-order perturbation theory gives
\begin{equation}
\delta\varepsilon_{\mathrm{D}}=\pm\varepsilon\frac{2X_{cv}^{2}-\hbar^{2}\Omega^{2}}{2X_{cv}^{2}+\hbar^{2}\Omega^{2}}.
\nonumber
\end{equation}
The correction is proportional to $\varepsilon$  and therefore vanishes at the original band crossing. Thus, the periodic field renormalizes the edge-state dispersion but does not open a gap at $\varepsilon=0$ and consequently does not drive a conventional metal-dielectric transition. 

The same perturbative treatment applied to the Floquet eigenstates gives the low-energy transverse matrix element
\begin{equation}
\left\langle x\right\rangle \approx x_{cv}\left[1-\frac{16\varepsilon^{2}\hbar^{2}\Omega^{2}X_{cv}^{2}}{\left(2X_{cv}^{2}+\hbar^{2}\Omega^{2}\right)^{3}}\right]+O(\varepsilon^4).
\nonumber
\end{equation}
Thus, the x-polarized transition remains finite at the zero-energy crossing, where $\left\langle x\right\rangle =x_{cv}$, and is only weakly modified away from it. In contrast, the longitudinal matrix element vanishes for the transition between the two central Floquet branches near $\varepsilon=0$. 

At finite energy, several $y$-polarized transitions are allowed among the six Floquet branches. Near the one-photon resonance, $2\varepsilon\approx\hbar\Omega$, the strongest by far are the symmetry-related transitions between the two branches forming the avoided crossing in each Floquet block. These arise from the resonant mixing of $(c_{-1},v_{0})$ in the upper block and $(c_{0},v_{1})$ in the lower block. Within the rotating-wave approximation, the corresponding longitudinal velocity matrix element is
\begin{equation}
\left\langle \hat{v}_{y}\right\rangle =\frac{\left|X_{cv}\right|\partial_{k_{y}}\varepsilon}{\hbar\sqrt{\left(\varepsilon-\frac{\hbar\Omega}{2}\right)^{2}+X_{cv}^{2}}}.
\nonumber
\end{equation}

It should be noted that the rotating-wave approximation is valid only near the one-photon resonance. Near $\varepsilon=0$, both neighboring Floquet replicas must be retained, since neglecting the counter-rotating contribution results in the opening of a gap at the zero-energy crossing.

\end{document}